\documentclass[%
 reprint,
 amsmath,amssymb,
 aps,
]{revtex4-1}

\usepackage{graphicx}
\usepackage{dcolumn}
\usepackage{bm}

\begin{document}

\preprint{APS/123-QED}

\title{Mesoscale modeling of colloidal suspensions with adsorbing solutes}

\author{Rei Tatsumi}
 \email{tatsumi@sogo.t.u-tokyo.ac.jp}
\author{Osamu Koike}%
\author{Yukio Yamaguchi}%
\affiliation{%
 Department of Chemical System Engineering, The University of Tokyo, Tokyo 113-8656, Japan
}%


\date{\today}

\begin{abstract}
We construct a mesoscale model of colloidal suspensions
that contain solutes reversibly adsorbing onto the colloidal particle surfaces.
The present model describes the coupled dynamics of the colloidal particles, the host fluid, and the solutes
through the Newton-Euler equations of motion, the hydrodynamic equations, and the advection-diffusion equation, respectively. 
The solute adsorption is modeled through a square-well potential,
which represents a short-range attractive interaction between a particle and a solute molecule. 
The present model is formulated to be solved through direct numerical simulations.
Some numerical results are presented to validate the simulations.
The present model enables investigations of solute adsorption effects 
in the presence of a fluid flow and an inhomogeneous solute concentration distribution.
\end{abstract}

\pacs{Valid PACS appear here}
\maketitle


\section{\label{s1}Introduction}
The dynamics of colloidal particles is appropriately investigated 
by means of mesoscale modeling. 
Mesoscale models explicitly describe the motion of the colloidal particles and the host fluid, 
thereby enabling exact evaluation of the hydrodynamic interactions among the particles 
mediated by the motion of the surrounding fluid~\cite{B1, B2}.
In many practical applications,
solutes, such as electrolytes, surfactants, and adsorbing polymers, are added 
to control the dispersion stability of colloidal suspensions~\cite{B1, B3}.
The mesoscale model can describe the influence of solutes on the dynamics of colloidal particles
by additionally considering the temporal evolution of the solute concentration field.
 
In the present study, 
we construct a mesoscale model of colloidal suspensions with an uncharged adsorbing solute such as a soluble polymer.
The solute molecules reversibly adsorb onto the particle surfaces to form thick adsorption layers, 
whose overlapping generates interparticle forces~\cite{B3, B4}.
We describe the solute adsorption through a square-well potential,
which represents a short-range attractive interaction between a particle and a solute molecule. 
This potential model reproduces the interparticle forces resulting from the overlapping of the adsorption layers, 
such as an attractive force induced by polymer bridging and a repulsive force induced by an increased osmotic pressure~\cite{B5, B6, B7}.
A decrease in the solute concentration around the particles, namely negative adsorption,
can be described by changing the particle-solute interaction to be repulsive,
thereby generating an attractive interparticle force referred to as a depletion force~\cite{B8}.
The present model also reproduces a particle motion caused by an inhomogeneity of solute concentration distribution,
such as diffusiophoresis~\cite{B9, B10}.
We describe the motion of the colloidal particles, the temporal evolution of the fluid flow field, and that of the solute concentration field
through the Newton-Euler equations of motion, the hydrodynamic equations, the advection-diffusion equation, respectively.
These governing equations of the present model are coupled each other,
an explicit form of which equations we derive.
The solute transport is contributed by the fluid flow as advective transport
and is affected by the particles through boundary conditions on the diffusion flux derived from 
the solute impermeability and adsorptivity to the particles.
The coupled dynamics of the particles and the fluid
is affected by the particle-solute adsorption interaction and the osmotic pressure in the adsorption layers.

The present model is formulated to be solved through direct numerical simulations,
of which scheme based on the immersed boundary method we have constructed ~\cite{B11, B12}.
In this method, the fluid flow field is also defined inside the particles,
where a body force is added to satisfy no-slip boundary conditions at the particle surfaces. 
To indicate the domain occupied by the particles, 
we introduce a function that is equal to unity inside the particles and zero outside them. 
The two domains are smoothly connected through a thin interfacial region with finite thickness 
to represent the particle-fluid interface on the discrete computational grid points.
In the present model, we also introduce another function to indicate the domain occupied by the adsorption layers. 
The boundary conditions on the diffusion flux 
are imposed by use of the indicator functions for the particles and the adsorption layers.

We examine the validity of the numerical simulations by comparing with some analytical solutions derived in this paper;
one is a steady-state solute concentration distribution around an isolated particle with imposing an external concentration gradient, 
and another is interparticle force generated by the adsorption layer overlapping.
We then investigate the influence of the solute adsorption on the steady-state sedimentation velocity of particles
as an example of the situation where the dynamics of the particles, the fluid, and the solute molecules are coupled.

\section{\label{s2}Model}
We model a colloidal suspension as a system in which rigid spherical particles with the same size and mass are dispersed in a Newtonian fluid.
The fluid contains one solute species;
the solute molecules reversibly adsorb onto the surface of the particles.
The reversible adsorption, namely physical adsorption, is described through a particle-solute interaction.
For the sake of simplicity, we ignore solute-solute and solute-solvent interactions;
in this assumption, the thermodynamic properties of the fluid is given by those of the ideal solution.  
 
In the present model, we consider the motion of the particles, the temporal evolution of the fluid flow field, and that of the solute concentration field.
The motion of the $i \mathchar`-$th particle is described by
the position $\boldsymbol{R}_i$, the translational velocity $\boldsymbol{V}_i$, and the angular velocity $\boldsymbol{\Omega}_i$.
With the assumption that the fluid is incompressible,
the fluid flow is described by the velocity field $\boldsymbol{v}(\boldsymbol{r}, t)$ and the pressure field $p(\boldsymbol{r}, t)$,
where $\boldsymbol{r}$ is the positional vector and $t$ is the time.
The solute concentration field $c(\boldsymbol{r}, t)$ is given as the number density of the solute molecules.

\subsection{\label{s2-1}Particle-solute interaction}

\begin{figure}[tbp]
\centering
\includegraphics[width=6cm, clip]{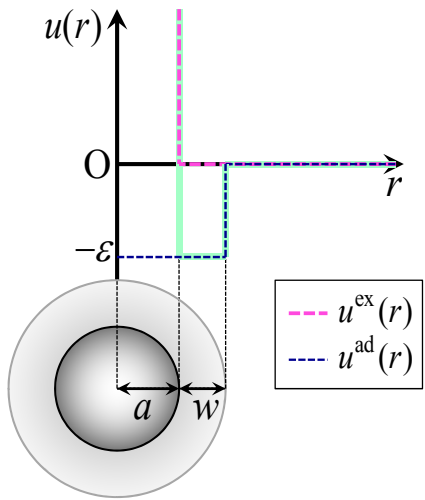} 
\caption{\label{f1} 
(Color online)
Interaction potential between a particle and a solute molecule.
The potential $u$ is described by a thick solid line 
and consists of the hard-core exclusion potential $u^\mathrm{ex}$ and the adsorption potential $u^\mathrm{ad}$,
which potentials are described by broken lines.  
}
\end{figure}

We describe the interaction between a particle and a solute molecule with a square-well potential,
which is expressed as follows:
\begin{eqnarray}
u(r) = u^\mathrm{ex}(r) + u^\mathrm{ad}(r)
\label{e2-1-1},
\end{eqnarray}
\begin{eqnarray}
u^\mathrm{ex}(r)
= \left\{ \begin{array}{ll} \vspace{6pt}
\infty, & 0 < r < a,\\ 
0, & r \geq a,\\ 
\end{array} \right.
\label{e2-1-2}
\end{eqnarray}
\begin{eqnarray}
u^\mathrm{ad}(r)
= \left\{ \begin{array}{ll} \vspace{6pt}
-\varepsilon, & 0 < r < b,\\
0, & r \geq b,\\ 
\end{array} \right.
\label{e2-1-3}
\end{eqnarray}
where $r$ represents the distance of the solute molecule from the center of the particle.
A sketch of the square-well potential is shown in Fig.~\ref{f1}.
The square-well potential $u$ consists of two contributions: the hard-core exclusion potential $u^\mathrm{ex}$ and the adsorption potential $u^\mathrm{ad}$.
The hard-core exclusion potential $u^\mathrm{ex}$ imposes the solute impermeability to the particle.
The size of the solute molecules is assumed to be negligibly smaller than that of the particles,
and hence the closest center-to-center distance between a particle and a solute molecule is equal to the particle radius $a$. 
The square-well potential is characterized by the well-width $w$ and the well-depth $\varepsilon$.
The adsorption layer is described as a spherical shell enclosed between two concentric spheres of radii $a$ and $b = a + w$.
The well-width and the well-depth correspond to the adsorption layer thickness and the adsorption energy, respectively.
Giving a negative value for the adsorption energy $\varepsilon$,
negative adsorption can be described.
We measure the adsorption energy by a non-dimensionalized value $\beta \varepsilon$,
where $\beta = (k_B T)^{-1}$, 
$k_B$ is the Boltzmann constant, and $T$ is the thermodynamic temperature.
Assuming the additivity of potentials,
the total potential acting on a solute molecule at the position $\boldsymbol{r}$ is given by
\begin{eqnarray}
U(\boldsymbol{r}) = \sum_i u( r_i )
\label{e2-1-4},
\end{eqnarray}
where $r_i = | \boldsymbol{r}_i |$, and $\boldsymbol{r}_i = \boldsymbol{r} - \boldsymbol{R}_i$.
We also define the total adsorption potential as
\begin{eqnarray}
U^\mathrm{ad}(\boldsymbol{r}) = \sum_i u^\mathrm{ad}( r_i )
\label{e2-1-5}.
\end{eqnarray}

\subsection{\label{s2-2}Indicator functions}

In the present formulation, 
fluid flow field is also defined inside the domain occupied by the particles.
The domain occupied by the $i \mathchar`-$th particle $P_i$ is given by
\begin{eqnarray}
P_i \equiv \{ \boldsymbol{r} \ | \ \chi_i(\boldsymbol{r}, t; a) = 1 \}
\label{e2-2-1},
\end{eqnarray}
where the function $\chi_i$ is a step function:
\begin{eqnarray}
\chi_i (\boldsymbol{r}, t; q)
= \left\{ \begin{array}{ll} \vspace{6pt}
1, & 0 < r_i < q,\\ 
0, & r_i \geq q.\\ 
\end{array} \right.
\label{e2-2-2}
\end{eqnarray}
We introduce a function $\Phi_i$ to indicate the domain $P_i$ as
\begin{eqnarray}
\Phi_i (\boldsymbol{r}, t) \equiv \chi_i ( \boldsymbol{r}, t; a)
\label{e2-2-3}.
\end{eqnarray}
A function indicating the union of all the domains $\{ P_i \}$ is also given by 
\begin{eqnarray}
\Phi (\boldsymbol{r}, t) \equiv \sum_i \Phi_i (\boldsymbol{r}, t)
\label{e2-2-4}.
\end{eqnarray}
We assume that the particles interact through a hard-core exclusion potential
to prevent them, namely the domains $\{ P_i \}$, from overlapping one another,
thereby limiting the range of the function $\Phi$ to $[0,1]$.
The particle velocity field $\boldsymbol{v}_p$, which represents the rigid motion of the particles,
is assigned to the velocity field inside the particles:
\begin{eqnarray}
\Phi \boldsymbol{v} = \Phi \boldsymbol{v}_p \equiv \sum_i \Phi_i (\boldsymbol{V}_i + \boldsymbol{\Omega}_i \times \boldsymbol{r}_i)
\label{e2-2-5}.
\end{eqnarray}
This formulation is equivalent to 
the imposition of no-slip boundary conditions on the velocity field at the particle surfaces. 

We also define the domain $W_i$ as one where the adsorption potential is induced by the $i \mathchar`-$th particle,
namely the domain inside the outer sphere composing the adsorption layer of the $i \mathchar`-$th particle:
\begin{eqnarray}
W_i \equiv \{ \boldsymbol{r} \ | \ \chi_i(\boldsymbol{r}, t; b) = 1 \}
\label{e2-2-6}.
\end{eqnarray}
Since the domain $W_i$ includes the domain $P_i$,
the adsorption layer is represented by the domain $W_i \backslash P_i$.
We define a function $\Xi_i$ to indicate the domain $W_i$
as the Boltzmann factor of the adsorption potential induced by the $i \mathchar`-$th particle:
\begin{eqnarray}
\Xi_i (\boldsymbol{r}, t) &\equiv& \exp(-\beta u_i^\mathrm{ad})
\label{e2-2-7}, \\
&=& \exp \left[ \beta \varepsilon \chi_i ( \boldsymbol{r}, t; b ) \right] 
\label{e2-2-8},
\end{eqnarray}
where $u^\mathrm{ad}_i \equiv u^\mathrm{ad} ( r_i )$.
A function indicating the union of all the domains $\{ W_i \}$ is given by
\begin{eqnarray}
\Xi (\boldsymbol{r}, t)
&\equiv& \exp( -\beta U^\mathrm{ad} )
\label{e2-2-9}, \\
&=& \exp \left[ \beta \varepsilon \sum_i \chi_i ( \boldsymbol{r}, t; b ) \right]
\nonumber, \\
&=& \prod_i \Xi_i (\boldsymbol{r}, t)
\label{e2-2-10}.
\end{eqnarray}
The domains $\{ W_i \}$ can overlap one another,
and hence the function $\Xi$ is equal to $\exp (n \beta \varepsilon)$,
where $n$ is the overlapping number of the domains $\{ W_i \}$.
The function $\Xi$ is unity outside the domains $\{ W_i \}$ where $n = 0$.

Under the ideal solution assumption,
the chemical potential difference between the solute and solvent, $\mu$, 
is given as
\begin{eqnarray}
\mu &=& k_B T \ln \frac{c}{c^{\circ\hspace{-0.51em}-}} + U
\label{e2-2-11}, 
\end{eqnarray}
where $c^{\circ\hspace{-0.51em}-}$ is the solute concentration in the standard state.
The concentration field in equilibrium with a fixed configuration of the particles is derived from the condition that $\mu$ is constant as
\begin{eqnarray}
c_\mathrm{eq} &=& c_0 \exp (-\beta U)
\label{e2-2-12}, \\
&=& (1 - \Phi) \Xi c_0
\label{e2-2-13},
\end{eqnarray}
where $c_0$ is the bulk solute concentration.
The indicator functions give the influence of the particles on the concentration field.
The solute concentration is zero inside the particles, namely $\{ P_i \}$, due to the solute impermeability to the particles 
and changes discontinuously by a factor of $\exp (n \beta \varepsilon)$ in the adsorption layers, namely $\{ W_i \backslash P_i \}$.
For an arbitrary solute concentration field $c (\boldsymbol{r}, t)$,
we define a virtual concentration field $c^{\ast} (\boldsymbol{r}, t)$ 
as a continuous concentration field from which the influence of the particles is eliminated:
\begin{eqnarray}
c = (1 - \Phi) \Xi c^{\ast}
\label{e2-2-14}.
\end{eqnarray}
The virtual concentration field is also defined inside the particles.

\subsection{\label{s2-3}Advection-diffusion equation}

The solute concentration field obeys the following equation 
that describes the conservation of solute mass:
\begin{eqnarray}
\frac{\partial c}{\partial t} + \boldsymbol{\nabla} \cdot (c \boldsymbol{v}_s)  = 0
\label{e2-3-1},
\end{eqnarray}
where $\boldsymbol{v}_s$ is the transfer velocity field of the solute.
The diffusion flux $\boldsymbol{J}$, which is defined as the relative mass flux to the barycentric fluid motion, 
is given in the linear order as~\cite{B13}
\begin{eqnarray}
\boldsymbol{J} &\equiv& c(\boldsymbol{v}_s - \boldsymbol{v})
\label{e2-3-2}, \\
&=& -\Gamma \boldsymbol{\nabla} \mu 
\label{e2-3-3},
\end{eqnarray}
where $\Gamma$ is the transport coefficient. 
Substituting Eq.~(\ref{e2-2-11}) into Eq.~(\ref{e2-3-3}) yields an expression of the diffusion flux as
\begin{eqnarray}
\boldsymbol{J} = - D (1 - \Phi) \Xi \boldsymbol{\nabla} c^{\ast}  
\label{e2-3-5},
\end{eqnarray}
where $D = \Gamma k_B T /c$ is the diffusion coefficient of the solute;
we assume a constant diffusion coefficient in the present model.
Equation~(\ref{e2-3-5}) indicates 
that the influence of the particles on the diffusion flux appears 
in the same manner as that on the solute concentration field described by Eq.~(\ref{e2-2-14}).
This formula is equivalent to 
the imposition of boundary conditions, which are derived from the solute impermeability and adsorptivity to the particles,
on the diffusion flux. 

We consider the temporal evolution of the virtual concentration field $c^{\ast}$ 
instead of the real concentration field $c$.
Equation~(\ref{e2-3-1}) is rewritten by substituting Eq.~(\ref{e2-2-14}) as
\begin{eqnarray}
\frac{\partial}{\partial t}\left[ (1-\Phi) \Xi c^{\ast} \right] 
+ \boldsymbol{\nabla} \cdot \left[ (1-\Phi) \Xi c^{\ast} \boldsymbol{v}_s \right]  = 0
\label{e2-3-6}.
\end{eqnarray}
Since there is no solute transport through the particle-fluid interface,
mass conservation of the solute virtually distributed inside the particles is described as
\begin{eqnarray}
\frac{\partial }{\partial t}(\Phi \Xi c^{\ast}) + \boldsymbol{\nabla} \cdot ( \Phi \Xi c^{\ast} \boldsymbol{v}_p)  = 0
\label{e2-3-7}.
\end{eqnarray}
Adding each side of Eqs.~(\ref{e2-3-6}) and (\ref{e2-3-7}) together yields the following equation:
\begin{eqnarray}
\frac{\partial }{\partial t}(\Xi c^{\ast}) + \boldsymbol{\nabla} \cdot (\Xi c^{\ast} \boldsymbol{v}_s^{\ast})  = 0
\label{e2-3-8},
\end{eqnarray}
where
\begin{eqnarray}
c^{\ast} \boldsymbol{v}_s^{\ast} = c^{\ast} \boldsymbol{v} + \boldsymbol{J}^{\ast}
\label{e2-3-9},
\end{eqnarray}
\begin{eqnarray}
\boldsymbol{J}^{\ast} = \Xi^{-1} \boldsymbol{J} = - D(1-\Phi) \boldsymbol{\nabla} c^{\ast}
\label{e2-3-10}.
\end{eqnarray}
Equation~(\ref{e2-3-8}) is rewritten as
\begin{eqnarray}
\Xi \left[ \frac{\partial c^{\ast}}{\partial t} + \boldsymbol{\nabla} \cdot (c^{\ast} \boldsymbol{v}_s^{\ast}) \right]
+ c^{\ast} \left( \frac{\partial \Xi}{\partial t} + \boldsymbol{v}_s^{\ast} \cdot \boldsymbol{\nabla} \Xi \right) = 0
\label{e2-3-11}.
\end{eqnarray}
The indicator function $\Xi_i$ obeys the following advection equation:
\begin{eqnarray}
\frac{\partial \Xi_i}{\partial t} + \boldsymbol{V}_i \cdot \boldsymbol{\nabla} \Xi_i = 0
\label{e2-3-12}.
\end{eqnarray}
The advection velocity, 
which is not contributed by the angular velocity of the particle due to the spherical symmetry of the adsorption potential $u^\mathrm{ad}$,
is equal to the translational velocity of the particle.
The second term in the left-hand side of Eq.~(\ref{e2-3-11}) is rewritten by use of Eq.~(\ref{e2-3-12}) as follows:
\begin{eqnarray}
\frac{\partial \Xi}{\partial t} + \boldsymbol{v}_s^{\ast} \cdot \boldsymbol{\nabla} \Xi
= \sum_i \frac{\Xi}{\Xi_i} \left( \frac{\partial \Xi_i}{\partial t} + \boldsymbol{v}_s^{\ast} \cdot \boldsymbol{\nabla} \Xi_i \right) 
\nonumber, \hspace{-16.5em}\\
&=& \sum_i \frac{\Xi}{\Xi_i} ( \boldsymbol{v}_s^{\ast} - \boldsymbol{V}_i ) \cdot \boldsymbol{\nabla} \Xi_i 
\nonumber, \\
&=& -\beta \Xi \sum_i ( \boldsymbol{v}_s^{\ast} - \boldsymbol{V}_i ) \cdot \boldsymbol{\nabla} u^\mathrm{ad}_i
\nonumber, \\
&=& -\beta \varepsilon \Xi \sum_i ( \boldsymbol{v}_s^{\ast} - \boldsymbol{V}_i ) \cdot \hat{\boldsymbol{r}}_i \delta(r_i - b)
\label{e2-3-13},
\end{eqnarray}
where $\hat{\boldsymbol{r}}_i = \boldsymbol{r}_i / r_i$.
Substituting Eq.~(\ref{e2-3-13}) into Eq.~(\ref{e2-3-11}) yields
the equation describing the temporal evolution of the virtual concentration field as
\begin{eqnarray}
\frac{\partial c^{\ast}}{\partial t} &+& \boldsymbol{\nabla} \cdot (c^{\ast} \boldsymbol{v}_s^{\ast})
\nonumber \\
&=& \beta \varepsilon c^{\ast} \sum_i ( \boldsymbol{v}_s^{\ast} - \boldsymbol{V}_i ) \cdot \hat{\boldsymbol{r}}_i \delta(r_i - b)
\label{e2-3-14}.
\end{eqnarray}
The right-hand side of Eq.~(\ref{e2-3-14}) represents 
that the inflow/outflow of the virtual concentration to/from the domain $W_i$, namely the adsorption/desorption of the solute,
results in the dissipation/production of the virtual concentration at the surface $\partial W_i$.
This dissipation/production occurs to satisfy mass conservation of the solute,
whose real concentration in the adsorption layer is given by
the multiplication of the virtual concentration with a factor of $\exp(n\beta \varepsilon)$.
The transport velocity of the virtual concentration through the surface $\partial W_i$,
which moves with the velocity $\boldsymbol{V}_i$, is given by $\boldsymbol{v}_s^{\ast} - \boldsymbol{V}_i $.

We consider thermal fluctuations,
which affect the mesoscale dynamics of colloidal particles,
through a random diffusion flux $\boldsymbol{\zeta}$ added to Eq.~(\ref{e2-3-5}).
The random diffusion flux is a vector-valued stochastic variable that satisfies the fluctuation-dissipation relation~\cite{B14}:
\begin{eqnarray}
\langle \zeta_{\alpha} (\boldsymbol{r}, t) \zeta_{\beta} (\boldsymbol{r}', t') \rangle
= 2 D c \delta_{\alpha \beta} \delta (\boldsymbol{r} - \boldsymbol{r}') \delta (t - t')
\label{e2-3-15}.
\end{eqnarray}

\subsection{\label{s2-4}Hydrodynamic  equations}

The hydrodynamic equations of the incompressible fluid
consist of the conservation equations of mass and momentum as
\begin{eqnarray}
\boldsymbol{\nabla} \cdot \boldsymbol{v} = 0
\label{e2-4-1},
\end{eqnarray}
\begin{eqnarray}
\rho \left( \frac{\partial \boldsymbol{v}}{\partial t} + \boldsymbol{v} \cdot \boldsymbol{\nabla} \boldsymbol{v} \right)
= \boldsymbol{\nabla} \cdot \boldsymbol{\sigma} + \boldsymbol{f}_S + \boldsymbol{f}_P
\label{e2-4-2},
\end{eqnarray}
where $\rho$ is the mass density of the fluid.
The stress tensor is given by
\begin{eqnarray}
\boldsymbol{\sigma} = -(p + \pi) \boldsymbol{I} 
+ \eta [\boldsymbol{\nabla} \boldsymbol{v} + (\boldsymbol{\nabla} \boldsymbol{v})^T] 
\label{e2-4-3},
\end{eqnarray}
where $\eta$ is the shear viscosity.
The osmotic pressure $\pi$,
which is the pressure change in the adsorption layers accompanied by the solute adsorption,
is given by the van't Hoff's formula in the ideal solution assumption as
\begin{eqnarray}
\pi = k_B T (\Xi - 1) c^{\ast}
\label{e2-4-4}.
\end{eqnarray}
The body force $\boldsymbol{f}_P$ is added 
to constrain the velocity field inside the particles to be equal to $\boldsymbol{v}_p$ given in Eq.~(\ref{e2-2-5}).
The body force $\boldsymbol{f}_S$ is exerted by the particle-solute adsorption interaction
and is expressed as
\begin{eqnarray}
\boldsymbol{f}_S &=& -c \boldsymbol{\nabla} U^\mathrm{ad}
\label{e2-4-5},  \\
&=& k_B T (1 - \Phi) c^{\ast} \boldsymbol{\nabla} \Xi
\label{e2-4-6}.
\end{eqnarray}
In Eq.~(\ref{e2-4-2}),
the effects of the solute adsorption appear through the chemical potential given by Eq.~(\ref{e2-2-11}),
of which the first term (the entropic contribution) and the second term (the energetic contribution) 
yield the osmotic pressure gradient $-\boldsymbol{\nabla} \pi$ and the body force $\boldsymbol{f}_S$, respectively.

Thermal fluctuations are also considered for the velocity field through a random stress tensor $\boldsymbol{s}$ added to Eq.~(\ref{e2-4-3}).
The random stress tensor is a tensor-valued stochastic variable satisfying the fluctuation-dissipation relation~\cite{B14}:
\begin{eqnarray}
\langle s_{\alpha \beta} (\boldsymbol{r}, t) s_{\mu \nu} (\boldsymbol{r}', t') \rangle \hspace{12.0em}
\nonumber \\
= 2 k_B T \eta (\delta_{\alpha \mu} \delta_{\beta \nu} + \delta_{\alpha \nu} \delta_{\beta \mu}) \delta (\boldsymbol{r} - \boldsymbol{r}') \delta (t - t')
\label{e2-4-7}.
\end{eqnarray}

\subsection{\label{s2-5}Newton-Euler equations of motion}

The motion of the colloidal particles is governed by the Newton-Euler equations of motion:
\begin{eqnarray}
M_i \frac{\mathrm{d}}{\mathrm{d} t} \boldsymbol{V}_i = \boldsymbol{F}^H_i + \boldsymbol{F}^C_i + \boldsymbol{F}^S_i,
\ \ \ \ \frac{\mathrm{d}}{\mathrm{d} t} \boldsymbol{R}_i = \boldsymbol{V}_i
\label{e2-5-1},
\end{eqnarray}
\begin{eqnarray}
\boldsymbol{I}_i \cdot \frac{\mathrm{d}}{\mathrm{d} t} \boldsymbol{\Omega}_i &=& \boldsymbol{N}^H_i
\label{e2-5-2}.
\end{eqnarray}
The $i \mathchar`-$th particle has the mass $M_i$ and the inertia tensor $\boldsymbol{I}_i$.
The force $\boldsymbol{F}^C_i$ is exerted by direct interactions among the particles and prevents the overlap of the particles.  
The hydrodynamic force $\boldsymbol{F}^H_i$ and torque $\boldsymbol{N}^H_i$ stem from the momentum exchange between the particle and fluid:
\begin{eqnarray}
\boldsymbol{F}^H_i = \int_{\partial P_i} \boldsymbol{\sigma} \cdot \mathrm{d} \boldsymbol{S}
\label{e2-5-3},
\end{eqnarray}
\begin{eqnarray}
\boldsymbol{N}^H_i = \int_{\partial P_i}  \boldsymbol{r}_i
\times (\boldsymbol{\sigma} \cdot \mathrm{d}\boldsymbol{S})
\label{e2-5-4},
\end{eqnarray}
where $\mathrm{d}\boldsymbol{S}$ is the area element vector. 
The integrations are performed over the surface of the particle $\partial P_i$.
The force $\boldsymbol{F}^S_i$ is the net force through the adsorption interaction between the particle and solute molecules:
\begin{eqnarray}
\boldsymbol{F}^S_i &=& \int c \boldsymbol{\nabla} u^\mathrm{ad}_i \mathrm{d} \boldsymbol{r}
\label{e2-5-5}, \\
&=& k_B T (e^{\beta \varepsilon} - 1) \int_{\partial W_i} \frac{\Xi}{\Xi_i} (1 - \Phi) c^{\ast} \mathrm{d} \boldsymbol{S}
\label{e2-5-6},
\end{eqnarray}
which describes the integration of the stress resulting from the discontinuous solute concentration change 
at the surface of the adsorption layer $\partial W_i$.
Since the direction of each force exerted by solute molecules, $\boldsymbol{\nabla} u^\mathrm{ad}_i$, is normal to the particle surface,
no torque is induced on the particle by the particle-solute adsorption interaction.

\section{\label{s3}Analytical solutions}

\subsection{\label{s3-1}Solute distribution disturbed by a particle}

Suppose that a constant solute concentration gradient $\boldsymbol{K}$ is imposed on a quiescent fluid,
thereby generating a linear concentration distribution given by $c = c_0 + \boldsymbol{K} \cdot \boldsymbol{r}$ in the absence of particles.
We derive a disturbed concentration distribution in steady state when a single particle is fixed in the fluid.
The origin of the coordinate system is set at the center of the particle,
and hence the unit vector $\hat{\boldsymbol{r}} = \boldsymbol{r}/r$ with $r = |\boldsymbol{r}|$
is normal to the surfaces of the particle and the adsorption layer.
In the present situation, a steady-state concentration field obeys
\begin{eqnarray}
\boldsymbol{\nabla} \cdot \boldsymbol{J} = 0
\label{e3-1-1},
\end{eqnarray}
with the boundary condition
\begin{eqnarray}
c^{\ast} \rightarrow c_0 + \boldsymbol{K} \cdot \boldsymbol{r} \ \ \ \ \mathrm{as} \ \ r \rightarrow \infty
\label{e3-1-2}.
\end{eqnarray}
Equation~(\ref{e3-1-1}) is reformulated to the Laplace's equation with the boundary conditions at the surfaces of the particle and the adsorption layer: 
\begin{eqnarray}
\boldsymbol{\nabla}^2 c^{\ast} = 0
\label{e3-1-3},
\end{eqnarray}
\begin{eqnarray}
(\hat{\boldsymbol{r}} \cdot \boldsymbol{\nabla} c^{\ast})_{r = a^+} = 0
\label{e3-1-4},
\end{eqnarray}
\begin{eqnarray}
(\hat{\boldsymbol{r}} \cdot \boldsymbol{\nabla} c^{\ast})_{r = b^+} 
= e^{\beta \varepsilon}(\hat{\boldsymbol{r}} \cdot \boldsymbol{\nabla} c^{\ast})_{r = b^-}
\label{e3-1-5},
\end{eqnarray}
\begin{eqnarray}
(\hat{\boldsymbol{r}} \times \boldsymbol{\nabla} c^{\ast})_{r = b^+} 
= (\hat{\boldsymbol{r}} \times \boldsymbol{\nabla} c^{\ast})_{r = b^-}
\label{e3-1-6}.
\end{eqnarray}
The conditions given by Eqs.~(\ref{e3-1-4}) and (\ref{e3-1-5}) are derived from the discontinuity of the diffusion flux described in Eq.~(\ref{e2-3-5}).
The condition given by Eq.~(\ref{e3-1-6}), which is derived from the irrotatinality of $\boldsymbol{\nabla} c^{\ast}$,
represents the continuity of the concentration gradient in the tangential directions to the surface of the adsorption layer.
The solution of Eqs.~(\ref{e3-1-2})-(\ref{e3-1-6}) is obtained as
\begin{eqnarray}
c^{\ast} - c_0
= \left\{ \begin{array}{ll} \vspace{8pt}
\displaystyle \alpha \left( 1 + \frac{a^3}{2r^3} \right) ( \boldsymbol{K} \cdot \boldsymbol{r} ), & a < r < b,\\
\displaystyle \left( 1 + \gamma \frac{b^3}{2r^3} \right) ( \boldsymbol{K} \cdot \boldsymbol{r} ), & r \geq b,\\ 
\end{array} \right.
\label{e3-1-7}
\end{eqnarray}
where
\begin{eqnarray}
\alpha = \frac{3}{A + 2B}, \ \  \gamma = \frac{2(-A + B)}{A + 2B}
\label{e3-1-8},
\end{eqnarray}
\begin{eqnarray}
A = e^{\beta \varepsilon} \left( 1 - \frac{a^3}{b^3} \right), \ \ B = 1 + \frac{a^3}{2b^3}
\label{e3-1-9}.
\end{eqnarray}

The solutions with the adsorption layer thickness $w/a = 0.5$ 
and some values of the adsorption energy $\beta \varepsilon$ are shown in Fig.~\ref{f4}.
The result of non-adsorbing solute ($\beta\varepsilon = 0$) indicates that
the solute diffusion from high to low concentration is hampered by the particle, 
thereby increasing and decreasing the solute concentration 
in front of and behind the particle in the direction of $\boldsymbol{K}$, respectively.
The disturbance is diminished/enhanced by the positive/negative solute adsorption.
This change reflects the influence of the solute adsorption on the diffusion flux around the particle described by Eq.~(\ref{e2-3-5}),
which indicates that the positive/negative adsorption increases/decreases 
the magnitude of the diffusion flux in the adsorption layer by a factor of $\exp (\beta \varepsilon)$.

\subsection{\label{s3-2}Interparticle force generated by adsorption layer ovelapping}

\begin{figure}[tbp]
\centering
\includegraphics[width=6cm, clip]{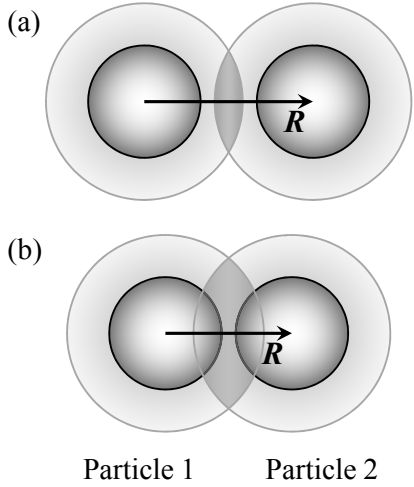}
\caption{\label{f2} 
Configuration of two particles.
The particles are situated in a fluid with a center-to-center distance $R = |\boldsymbol{R}|$.
(a) The adsorption layers overlap each other without one particle facing to the adsorption layer of the other as $a+b \leq R < 2b$.
(b) One particle faces to the adsorption layer of the other as $R \leq a+b$.
}
\end{figure}

Suppose two particles arranged in a fluid as described in Fig.~\ref{f2}.
The center-to-center vector from the particles 1 to 2 is given by
$\boldsymbol{R} = \boldsymbol{R}_2 - \boldsymbol{R}_1$,
the unit vector of which direction is expressed as $\hat{\boldsymbol{R}} = \boldsymbol{R}/R$ with $R = |\boldsymbol{R}|$.
We consider an equilibrium solute concentration distribution with fixing the positions of the particles.
When each particle is isolated from the other,
zero net force acts on each particle because of the spherically symmetric solute concentration distribution around the particle.
On the other hand, when the adsorption layers of the particles overlap each other, 
the spherical symmetry of the solute concentration distribution is broken,
and hence non-zero net force acts on each particle.
The forces acting on the particle 2, 
namely the force induced by the particle-solute adsorption interaction $\boldsymbol{F}^S_2$ and that induced by the osmotic pressure $\boldsymbol{F}^H_2$,
are respectively evaluated by Eqs.~(\ref{e2-5-6}) and (\ref{e2-5-3}) as
\begin{eqnarray}
\boldsymbol{F}^S_2 
&=& k_B T c_0 (e^{\beta \varepsilon} - 1) \int_{\partial W_2} \Xi_1 (1 - \Phi) \mathrm{d} \boldsymbol{S}
\nonumber, \\
&=& -k_B T c_0 (e^{\beta \varepsilon} - 1) 
\nonumber \\
&& \times \left[ e^{\beta \varepsilon} (S_{WW} - S_{PW}) - S_{WW} \right] \hat{\boldsymbol{R}}
\label{e3-2-1}, 
\end{eqnarray}
\begin{eqnarray}
\boldsymbol{F}^H_2 &=& - \int_{\partial P_2} \pi \mathrm{d} \boldsymbol{S}
\label{e3-2-2}, \\
&=& - k_B T c_0 e^{\beta \varepsilon} \int_{\partial P_2} \Xi_1 \mathrm{d} \boldsymbol{S}
\nonumber, \\
&=& k_B T c_0 (e^{2 \beta \varepsilon} - e^{\beta \varepsilon}) S_{PW} \hat{\boldsymbol{R}}
\label{e3-2-3},
\end{eqnarray}
where ${S}_{WW}$ and ${S}_{PW}$ are the circular areas 
enclosed by the edges of the surfaces $\partial W_2 \cap W_1$ and $\partial P_2 \cap W_1$ (or $\partial W_2 \cap P_1$), respectively.
In the derivation of Eqs.~(\ref{e3-2-1}) and (\ref{e3-2-3}),
we employ the following relations derived from the axisymmetry about the $\hat{\boldsymbol{R}}$ axis:
\begin{eqnarray}
\int_{\partial W_2 \cap W_1} \hspace{-0.6em} \mathrm{d} \boldsymbol{S} = S_{WW} \hat{\boldsymbol{R}}
\label{e3-2-4},
\end{eqnarray}
\begin{eqnarray}
\int_{\partial P_2 \cap W_1} \hspace{-0.6em} \mathrm{d} \boldsymbol{S} = \int_{\partial W_2 \cap P_1} \hspace{-0.6em} \mathrm{d} \boldsymbol{S}
= S_{PW} \hat{\boldsymbol{R}}
\label{e3-2-5}.
\end{eqnarray}
The areas ${S}_{WW}$ and ${S}_{PW}$ are expressed as functions of the center-to-center distance between the particles $R$:
\begin{eqnarray}
S_{WW} = \frac{\pi}{4} (4 b^2 - R^2) \ \ \ \ \mathrm{as} \ \ R \leq 2b
\label{e3-2-6},
\end{eqnarray}
\begin{eqnarray}
S_{PW} = \frac{\pi}{4 R^2} \left[ 4 b^2 R^2 - (R^2 + b^2 - a^2)^2 \right] \hspace{2.5em}
\nonumber \\
\mathrm{as} \ \ R \leq a+b
\label{e3-2-7}.
\end{eqnarray}
The net force acting on the particle 2 is given by
\begin{eqnarray}
\boldsymbol{F}_2 &=& \boldsymbol{F}^H_2 + \boldsymbol{F}^S_2
\nonumber \\ 
&=& k_B T c_0 (e^{\beta \varepsilon} - 1)
\nonumber \\
&& \times \left[ 2 e^{\beta \varepsilon} S_{PW} - (e^{\beta \varepsilon} - 1) S_{WW} \right] \hat{\boldsymbol{R}}
\label{e3-2-8},
\end{eqnarray}
which is the same as the result obtained in Ref.~\cite{B6}.

\begin{figure}[tbp]
\centering
\includegraphics[width=7cm]{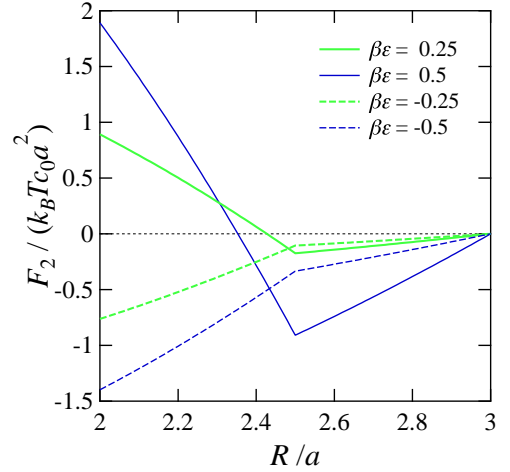}
\caption{\label{f3}
(Color online)
Force acting on a particle, whose adsorption layer overlaps with that of another particle,
as a function of the center-to-center distance between the particles for various adsorption energies $\beta \varepsilon$.
The adsorption layer thickness is set as $w/a = 0.5$.
The negative and positive values correspond to attractive and repulsive forces, respectively.
}
\end{figure}

Figure~\ref{f3} shows the force acting on the particle 2 for the adsorption energies $\beta \varepsilon = \pm0.25, \pm0.5$
with the adsorption layer thickness $w/a = 0.5$.
We first consider the positive adsorption described by $\beta\varepsilon > 0$.
The overlapping of the adsorption layers as $R \leq 2b$ (where $2b = 3$) leads to 
an increased solute concentration at the overlapped adsorption layer, 
toward which the particles are attracted through the particle-solute adsorption interaction (Fig.~\ref{f2}a). 
In other words, an attractive force given by $\boldsymbol{F}^S_2$ acts on the particle, 
which force describes an attractive force induced by polymer bridging.
When two particles get closer as $R \leq a+b$ (where $a+b = 2.5$), 
one particle faces to the adsorption layer of the other 
and goes through a high osmotic pressure at the overlapped adsorption layer (Fig.~\ref{f2}b), 
thereby imposing a repulsive force given by $\boldsymbol{F}^H_2$ on the particle. 
Consequently, there exists an equilibrium interparticle distance as $R/a  \approx 2.43$ and 2.35 
for $\beta\varepsilon = 0.25$ and 0.5, respectively;
the particles will assemble with some interparticle separation.
On the other hand, when we consider the negative adsorption described by $\beta\varepsilon < 0$, 
a decreased solute concentration at the overlapped negative adsorption layers, 
where both of the particle-solute repulsive force and the osmotic pressure is diminished,
imposes a totally attractive interparticle force.
This attractive force is referred to as a depletion force~\cite{B8}.

\section{\label{s4}Numerical results}

Direct numerical simulations of the present model are performed 
by use of the scheme given in our previous papers~\cite{B11, B12}.
To represent the indicator functions on the discrete computational grids,
we use a continuous function as a substitute for the step function given by Eq.~(\ref{e2-2-2})~\cite{B15}:
\begin{eqnarray}
\chi_i (\boldsymbol{r}; q) = \frac{f[(q + \xi/2) - r_i]}{f[(q + \xi/2) - r_i] + f[r_i - (q - \xi/2)]}
\label{e4-1},
\end{eqnarray}
\begin{eqnarray}
f(x)
= \left\{ \begin{array}{ll} \vspace{6pt}
\exp(- \Delta^2 /x^2), & x \geq 0,\\ 
0, & x < 0,\\ 
\end{array} \right.
\label{e4-2}
\end{eqnarray}
where $\Delta$ is the spacing of simulation grids.
The function given by Eq.~(\ref{e4-1}) represents that
the domains inside ($\Phi = 1$) and outside ($\Phi = 0$) the particles are smoothly connected by a thin interface of thickness $\xi$,
which is set as $\xi/\Delta = 2$ in all of the simulations in the present study.
In this description,
the particle-fluid interface has a finite volume and thereby is supported by multiple grid points.

The equations of motion of particles given by Eqs.~(\ref{e2-5-1}) and (\ref{e2-5-2}) are solved with the Euler scheme; 
time evolution of the particle position is solved with the Crank-Nicolson scheme.  
The hydrodynamic equations given by Eqs.~(\ref{e2-4-1}) and (\ref{e2-4-2}) 
are solved using the SIMPLEST (semi-implicit method for pressure-linked equations shortened)~\cite{B16}.
The detailed expression of the constraint body force $\boldsymbol{f}_p$ is described in the previous paper~\cite{B12}.
To solve the diffusion-advection equation of the virtual concentration field given by Eq.~(\ref{e2-3-14}), 
the flux term in the left-hand side is rewritten by use of the solenoidal condition Eq.~(\ref{e2-4-1}) as
\begin{eqnarray}
\boldsymbol{\nabla} \cdot (c^{\ast} \boldsymbol{v}^{\ast}_s)
= \boldsymbol{v} \cdot \boldsymbol{\nabla} c^{\ast} + \boldsymbol{\nabla} \cdot \boldsymbol{J}^{\ast}
\label{e4-3}.
\end{eqnarray}
The first and second terms in the right-hand side represent the advection and the diffusion of the solute, respectively.
The discretization in space of Eq.~(\ref{e2-3-14}) is carried out;
a second-order central differencing is employd for the diffusion term,
while a third-order upwind differencing is employed for the advection term and the right-hand side of the equation.
The time integration is performed by the second-order Runge-Kutta scheme (the Heun scheme).
In the following simulations, we neglect the thermal fluctuations introduced through Eqs.~(\ref{e2-3-15}) and (\ref{e2-4-7}).

\subsection{\label{s4-1}Solute distribution disturbed by a particle}

\begin{figure}[tbp]
\centering
\includegraphics[width=7cm]{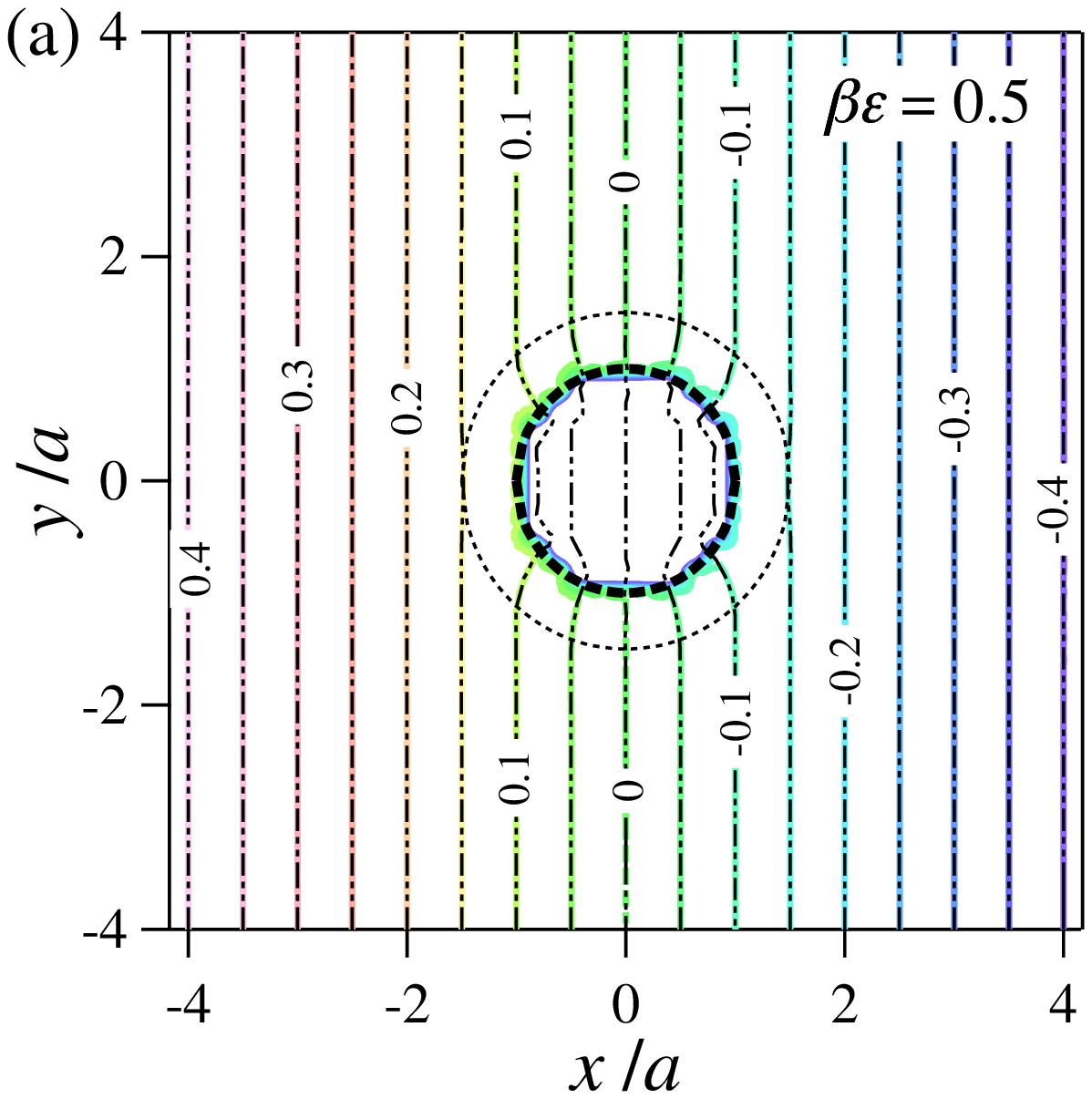} 
\includegraphics[width=7cm]{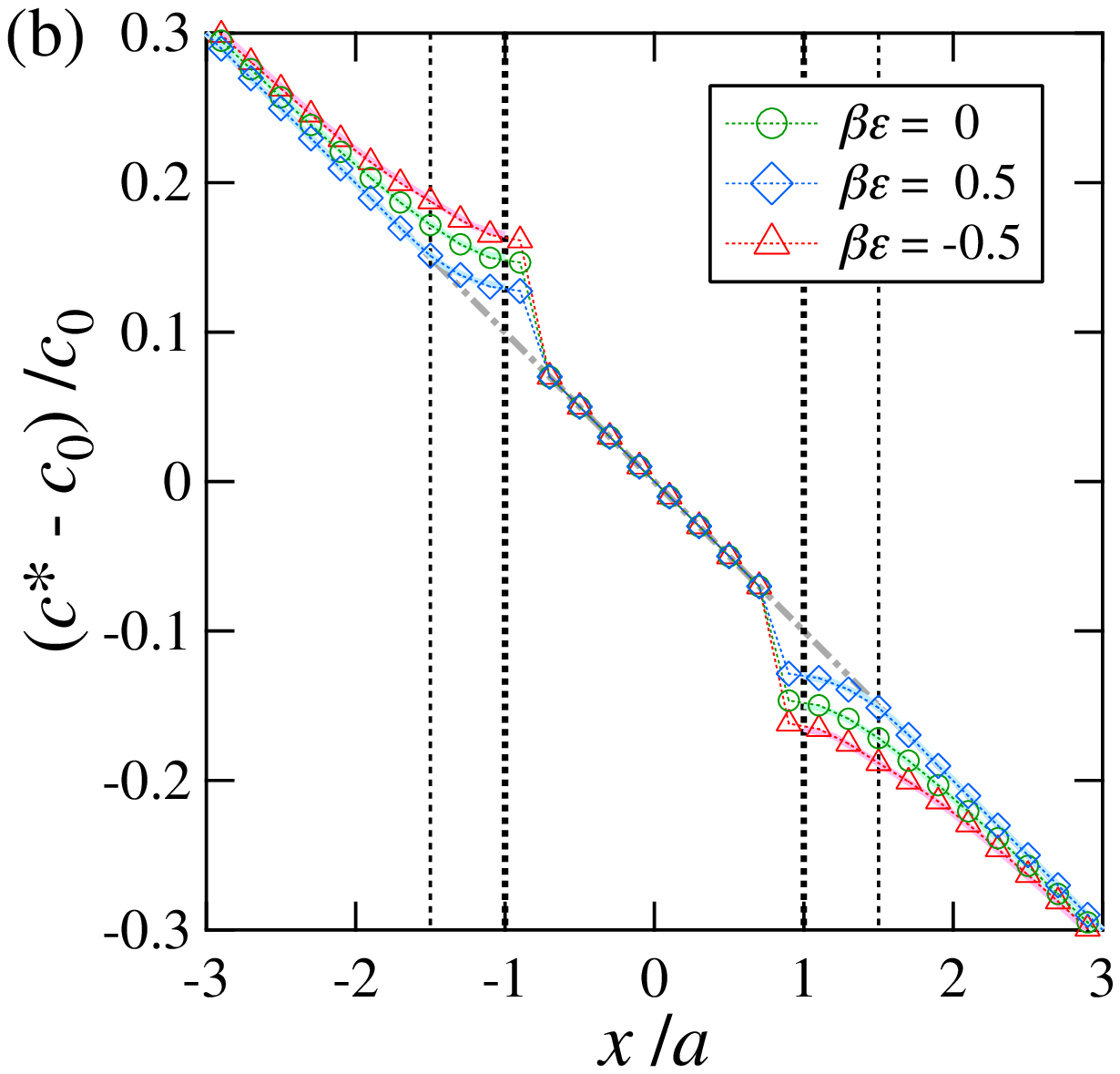} \vspace{-0.5em}
\caption{\label{f4}
(Color online)
Virtual solute concentration field $c^{\ast}$ around a particle 
with the imposed concentration gradient $\boldsymbol{K} = -0.1(c_0/a) \hat{\boldsymbol{x}}$.
The adsorption layer thickness is $w/a = 0.5$.
(a) Contour lines of the concentration field at the plane of $z = 0$ for $\beta \varepsilon = 0.5$.
The color lines represent the analytical solution,
while the black dashed-two dotted lines represent the simulation result.
The surfaces of the particle and the adsorption layer are described by thick and thin broken lines, respectively.
(b) Concentration distribution on the line of $y = z = 0$.
The adsorption energy takes the values of $\beta \varepsilon = 0$, $\pm 0.5$.
The solid lines represent analytical solutions.
The dashed double-dotted line represents the concentration distribution in the absence of the particle, $c^{\ast} = |\boldsymbol{K}| x$.
The thick and thin broken lines represent the positions where surfaces of the particle and the adsorption layer are present, respectively.
}
\end{figure}

To validate the numerical simulation of the advection-diffusion equation given by Eq.~(\ref{e2-3-14}), 
we calculated a solute concentration distribution around a particle with imposing a concentration gradient.
The analytical solution is derived in $\S$\ref{s3-1}.
The simulation box is set as a cube of side length $L = 10a$.
A single particle is fixed on the center of the simulation box.
We introduce a Cartesian coordinate system with the origin at the center of the simulation box and 
the axes parallel to the sides of the simulation box.
The imposed concentration gradient is $\boldsymbol{K} = -0.1(c_0/a) \hat{\boldsymbol{x}}$,
where $\hat{\boldsymbol{x}}$ is the unit vector of $x$ direction.
The concentration at the boundary surfaces of $x = \pm L/2$ is fixed as $c_0 \mp |\boldsymbol{K}|L/2$;
periodic boundary conditions are imposed on the $y$ and $z$ directions.
 
The virtual concentration field $c^{\ast}$ around the particle in steady state is shown in Fig.~\ref{f4}.
The adsorption layer thickness and the adsorption energy take the values of $w/a = 0.5$ and $\beta \varepsilon = 0, \pm 0.5$, respectively.
Excellent agreement with the analytical solutions is obtained.
Influence from the periodic boundary conditions is slight
because the order ratio of the disturbance caused by the particle to the non-disturbed concentration field 
is proportional to $r^{-3}$ as described by Eq.~(\ref{e3-1-7}).

\subsection{\label{s4-2}Interparticle force generated by adsorption layer ovelapping}

\begin{figure}[tbp]
\centering
\includegraphics[width=7cm]{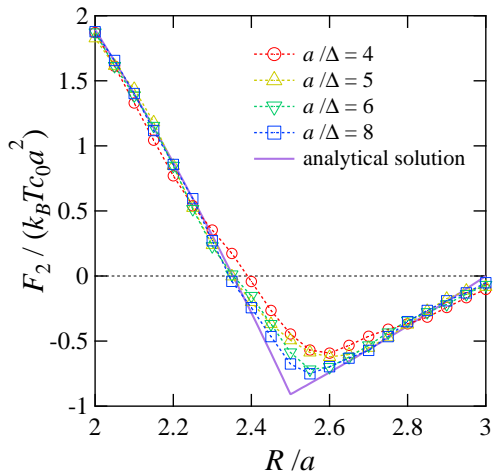} \vspace{-0.5em}
\caption{\label{f5}
(Color online)
Simulation estimation of force acting on a particle induced by the adsorption layer overlapping
with various grid resolutions of the particle $a/\Delta$.
The adsorption layer thickness and adsorption energy are $w/a = 0.5$ and $\beta \varepsilon = 0.5$, respectively.
}
\end{figure}

To validate the numerical evaluation of force induced on the particles by the solute adsorption,
we calculated the interparticle force between two particles generated by the adsorption layer overlapping.
The analytical solution is derived in $\S$\ref{s3-2}.
Figure~\ref{f5} shows the simulation results with various grid resolutions of the particle $a/\Delta$.
The adsorption layer thickness and the adsorption energy are set as $w/a = 0.5$ and $\beta \varepsilon = 0.5$, respectively.
A remarkable deviation from the analytical solution is observed around the kink at the minimum point.
This discrepancy is attributed to the smoothed particle-fluid interface in the present simulation.
An increase in the grid resolution $a/\Delta$
leads to the reduction in the deviation of the simulation result from the analytical solution.

\subsection{\label{s4-3}Solute adsorption effects in sedimentation}

\begin{figure}[tbp]
\centering
\includegraphics[width=7cm]{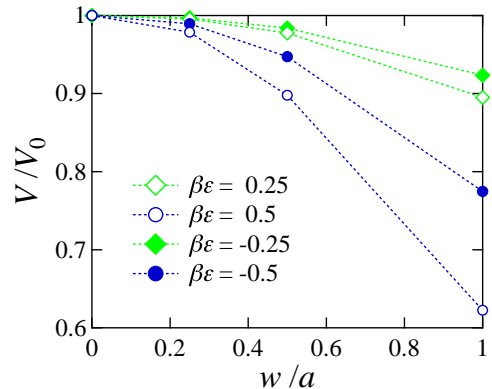} 
\caption{\label{f6} 
(Color online)
Sedimentation velocity of a particle with different adsorption energies $\beta \varepsilon$ 
as a function of the adsorption layer thickness $w/a$.
The velocity is normalized by the sedimentation velocity with no solute adsorption ($\beta \varepsilon = 0$), 
which is given by $V_0$.
The adsorption energy takes the values of $\beta \varepsilon = \pm0.25, \pm0.5$.
}
\end{figure}

\begin{figure}[tbp]
\centering
\includegraphics[width=5.3cm, clip]{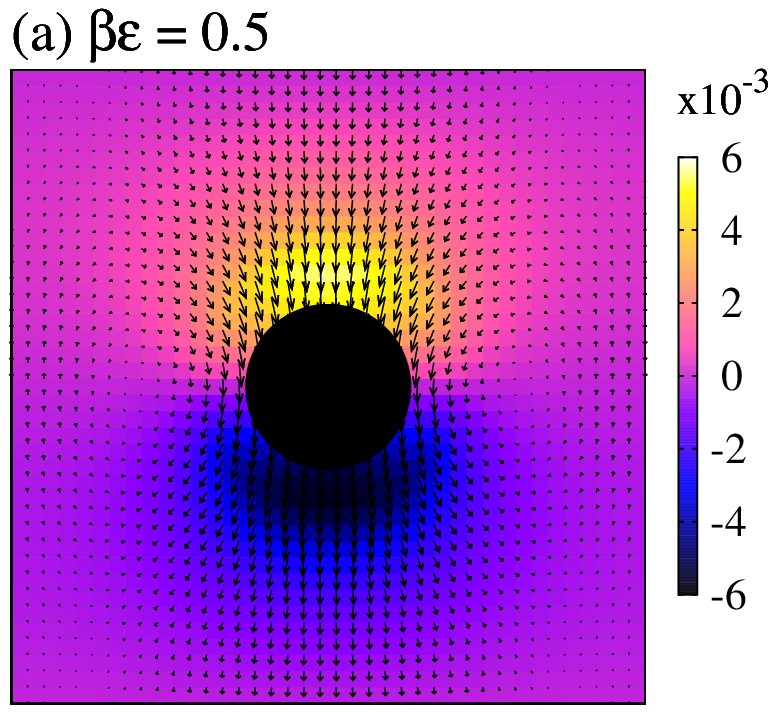} 
\includegraphics[width=5.3cm, clip]{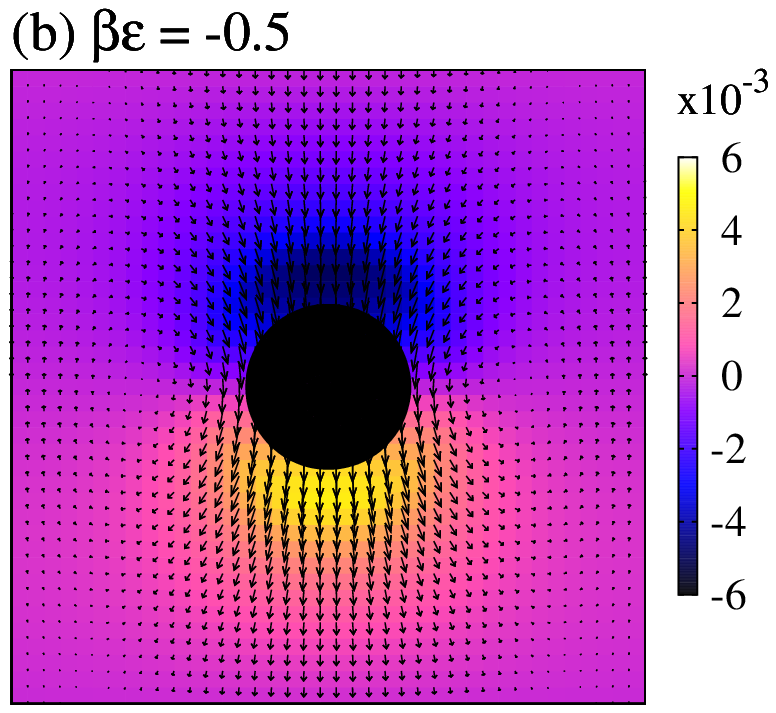} 
\caption{\label{f7} 
(Color online)
Deviation of virtual solute concentration $c^{\ast} - c_0$ and velocity field $\boldsymbol{v}$ around a particle, 
with the adsorption layer thickness $w/a = 0.5$.
The deviation of the virtual concentration is scaled by the bulk concentration $c_0$.
The adsorption energy takes the values of (a) $\beta \varepsilon = 0.5$ and (b) $\beta \varepsilon = -0.5$.
The cross-sections presented here are parallel to the particle motion direction and include the particle center.
The particle is represented by a black circle.
The direction of the particle velocity is to the downward in the pictures.
The deviation of the virtual solute concentration is described by a color scale,
which goes from negative (darker) to positive (lighter) deviation.
}
\end{figure}

We present the simulation results of a steady-state sedimentation velocity of particles arranged in a periodic array with an adsorbing solute;
in this situation, coupled dynamics of the particles, the fluid, and the solute is to be considered.
The simulation box is set as a cube of side length $L = 8a$ with periodic boundary conditions,
thereby forming a periodic array of the particles.
In the reduced units as $\Delta = \rho = \eta = 1$,
the bulk solute concentration is set as $c_0 k_B T = 0.0244$,
and the diffusion coefficient is set as $D = 0.1$ (the Schmidt number is $\mathrm{Sc} = \eta/\rho D = 10$).
The particles are continuously subjected to a constant force,
of which magnitude is set such that the particle Reynolds number with no solute adsorption ($\beta \varepsilon = 0$) 
is $\mathrm{Re}_p = \rho a V_0/\eta = 8 \times 10^{-3}$.

The steady-state sedimentation velocity of the particles as a function of the adsorption layer thickness is shown in Fig.~\ref{f6}.
The adsorption energy takes the values of $\beta \varepsilon = \pm0.25, \pm0.5$.
The sedimentation velocity is reduced with the increase in the adsorption layer thickness for both positive and negative adsorption,
and hence the particle Reynolds number is less than $8 \times 10^{-3}$ in all the cases.
The virtual concentration distribution and velocity field around one of the settling particles 
with $w/a = 0.5$ are depicted in Fig.~\ref{f7}.
The positive adsorption described by $\beta\varepsilon = 0.5$ (Fig.~\ref{f7}a) leads to 
a decreased and increased virtual concentration in front of and behind the particle in the direction of the particle motion, respectively,
while the negative adsorption described by $\beta\varepsilon = -0.5$ (Fig.~\ref{f7}b) leads to an opposite change in the virtual concentration.
In both cases,
the generated asymmetric concentration distribution induces a force on the particle in the direction opposite to the particle motion,
thereby reducing the sedimentation velocity.
The larger force induced on the particle according to the thicker adsorption layer is 
due to the larger surface area of the adsorption layer where solute molecules exert force on the particle.
The generation of the asymmetric concentration distribution can be explained in terms of Eq.~(\ref{e2-3-14});
the solute transported slowly relative to the particle motion is adsorbed and desorbed in front of and behind the particle.
Similar phenomenon has been investigated for charged particles in an electrolyte solution~\cite{B17, B18}.

\section{\label{s5}Conclusion}

We constructed a mesoscale model of colloidal suspensions containing an adsorbing solute.
We derived equations describing the coupled dynamics of the colloidal particles, the host fluid, and the solute molecules;
these equations were formulated to be solved through direct numerical simulations.
We performed the validation of the simulations
by comparing with the analytical solutions derived in the present paper for two situations: 
a steady-state solute concentration distribution around a particle with imposing an external concentration gradient 
and interparticle force generated by the adsorption layer overlapping.
We then computed the steady-state sedimentation velocity of particles arranged in a periodic array.
The reduction of the sedimentation velocity was observed according to an asymmetric solute concentration distribution around the particles,
which asymmetric distribution was generated by the solute adsorption and desorption at the surfaces of the adsorption layers.

The present model is appropriate to investigate solute adsorption effects on the dynamics of colloidal particles
in the presence of a fluid flow and an inhomogeneous solute concentration distribution.
As an interesting application of the present model,
we will be able to investigate the structure formation of colloidal particles
in a drying process of colloidal suspensions containing adsorbing solutes.
In this case, the present model is required to be developed
to consider the gas-liquid free surface motion according to evaporation~\cite{B19},
beneath which surface a high concentration layer of solutes can appear.


\nocite{*}

\bibliography{apssamp}

\end{document}